\documentclass[review, 12pt]{elsarticle}

\usepackage{graphicx}
\usepackage{amssymb}
\usepackage{amsthm}
\usepackage{amsmath}
\usepackage[latin9]{inputenc}
\usepackage{color}
\usepackage{blindtext}
\usepackage{lineno}
\usepackage{gensymb}

\usepackage{fleqn}
\usepackage{geometry}
\usepackage{natbib}

\newcommand{\lmp}{LiMnPO\(_4\)}
\newcommand{\lfp}{LiFePO\(_4\)}
\newcommand{\lmmp}{Li\(M\)PO\(_4\)\ \(M\)=(Mn, Fe, Co, Ni) }
\newcommand{\lmfp}{Li(Mn,Fe)PO\(_4\)}
\newcommand{\lmfpx}{LiMn\(_{1-x}\)Fe\(_x\)PO\(_4\)}
\newcommand{\lmfuenf}{LiMn\(_{0.5}\)Fe\(_{0.5}\)PO\(_4\)}

\newcommand{\vsli}{vs. Li/Li$^{+}$}
\newcommand{\etal}{\textsl{et~al.}}

\begin{document}

\begin{frontmatter}

\title{Anisotropic ionic conductivity of LiMn$_{1-x}$Fe$_x$PO$_4$ ($0\leq x \leq 1$) single crystals}

\author[1,3]{C.~Neef}
\author[1]{A.~Reiser}
\author[1]{E.~Thauer}
\author[1,2]{R.~Klingeler}
\ead{r.klingeler@kip.uni-heidelberg.de}

\address[1]{Kirchhoff Institute of Physics, Heidelberg University, D-69120 Heidelberg, Germany}
\address[2]{Centre for Advanced Materials, Heidelberg University, D-69120 Heidelberg, Germany}
\address[3]{Fraunhofer Institute for Systems and Innovation Research ISI, D-76139 Karlsruhe, Germany}

\begin{abstract}

We report AC-impedance studies on a series of high-quality LiMn$_{1-x}$Fe$_x$PO$_4$ single crystals with $0\leq x \leq 1$. Our results confirm quasi-one-dimensional transport in \lfp\ with fast Li-diffusion along the $b$-axis. The conductivities along the crystallographic $b$-, $c$- and $a$-axis differ by a factor of about 10, respectively. Whereas, the activation energy $E_{\rm A}$ of the effective diffusion process is particularly large for the $b$-axis and smallest for the $a$-axis. Remarkably, the $b$-axis ionic bulk conductivity of \lmfuenf\ is of the same order of magnitude as in undoped \lfp , which implies similarly fast Li-transport even upon 50~\% Mn-doping which, owing to the higher redox potential of the Mn$^{3+}$/Mn$^{2+}$-couple, yields enhanced energy density in lithium-ion batteries. The overall results of our impedance studies draw a far more complex picture than it would be expected from a simple one-dimensional ionic conductor. Our results suggest a strong contribution of crystal defects in real materials.

\end{abstract}


\end{frontmatter}


\section{Introduction}

Electronic and ionic conductivity are crucial parameters governing the performance of battery materials, including the charge and discharge rates, cycling stability and practically accessible capacity. Sufficient conduction is still one of the main challenges for enhancing Lithium-ion batteries. Accordingly, for usage in high-power applications, a cathode material like olivine-structured \lfp\ which is a good insulator at room temperature ($\sigma$ $\approx$ $10^{-9}$\,S\,cm$^{-1}$~\cite{Wang2007b}) has to be appropriately modified by nanoscaling in combination with carbon coating before displaying sufficient electrochemical activity~\cite{Wang2007b,JWang2012}. However, despite the great technological importance of Li-battery materials and of \lfp\ in particular, many of the fundamental mechanisms determining Li-intercalation remain to be fully understood. This includes the detailed relation between electronic and ionic conductivity and their microscopic mechanism. In addition, modelling of battery materials and of cell performances crucially depend on input parameters. In particular, theoretical predictions of improved design of battery materials down to the size and shape of primary particles as well as the hierarchical architecture of carbon and active materials requires precise knowledge of not only effective conductivities but validates parameters for the different crystallographic directions. In addition, knowledge of those parameters allows tailoring the morphology of the micro- and nanosized powders used in actual devices and to possibly improve the performance of Li-ion batteries by, e.g., utilizing oriented powders.

Ionic conductivity of olivine-structured phosphates has been particularly studied for \lfp\ both by experimental~\cite{Amin2008,Amin2007,Amin2008I,Li2008,Prabu2011} and theoretical methods~\cite{Gardiner2010,Adams2010,Yang2011}. Theoretical models suggest ionic diffusion along the crystallographic $b$-axis which implies rather one-dimensional (1D) Li-motion. Experimentally, however, there is no consensus on the transport as studies on single crystals provide qualitatively contradicting results. While Ref.~\cite{Li2008} reports Li-diffusion along the $b$-axis about 2 to 3 order of magnitude larger than along the $a$- and $c$-directions, measurements in Ref.~\cite{Amin2008,Amin2008II} imply rather two-dimensional conductivity in the ($b$,$c$)-plane. In real crystals and at finite temperature, various structural defects are to be expected which in particular in the case of 1D diffusion may strongly affect the actual conductivity~\cite{Malik2010}. Numerical studies on the formation of various Frenkel- and Schottky-defects show that in particular Li-Fe intersite defects are to be expected.~\cite{Gardiner2010} They are associated with site exchange between Li-ions at regular Li-positions $Li^{\times}_{Li}$ and corresponding transition metal ions positions $M^{\times}_M$ which may be described by $Li^{\times}_{Li} + M^{\times}_M \rightarrow Li'_M + M^{\cdot}_{Li}$. Depending on the synthesis route and thermal treatments, intersite exchange of a few percent can be expected.~\cite{Chung2008} Note, that in \lmfpx , the formation energy of intersite defects decreases with increasing $x$.~\cite{Gardiner2010} This is tentatively confirmed by previous structure analyses of the single crystals studied in the present work~\cite{Neef2017} which suggest about 1~\% of such defects in \lmp\ and 2~\% in \lfp .


In \lmmp , the cell voltage and the accordingly energy density are increasing with the actual redox potential which is enhanced when replacing Fe by Mn, Co, and Ni. In this respect, owing to its redox potential of 4.1~\vsli\ as compared to 3.4~V in \lfp , \lmp\ is a competitive cathode material offering about 20~\% higher energy density. Despite its much lower electrical conductivity carbon-coated nanostructured LiMnPO4 has been found to show a stable reversible capacity of up to 145~mAh\,g$^{-1}$.~\cite{Wang2009,Aravindan} M-site co-doping, even at relatively low Fe-levels of ~20~\%, enhances conductivity so that significantly higher rate capability and larger reversible capacity than the Mn-end member is realized.~\cite{Martha2009,Oh2012}

We report anisotropic ionic conductivity of a series of LiMn$_{1-x}$Fe$_x$PO$_4$ single crystals with doping levels 0 $\leq x \leq$ 1 grown at high Ar pressure. The results on \lfp\ confirm rather 1D ionic diffusion. Studies on crystals with $x\neq 1$ have not yet been reported. The end member \lmp\ displays very low conductivities. Remarkably, the mixed compounds \lmfp\ display moderate suppression of diffusion upon Mn-doping. Specifically, ionic bulk conductivity along the fast diffusion $b$-axis in \lmfuenf\ is of the same order of magnitude as in undoped \lfp .

\section{Experimental}

Single crystals of the compositions \lmfpx , with $x$ = 0, 0.3, 0.5, 1, were grown by the optical floating zone technique as described in Ref.~\cite{Neef2017} in detail. The samples were oriented by Laue-X-ray-diffraction perpendicular to the crystallographic main axes $a$, $b$, and $c$. Platelets with a thickness of 0.3\,mm and a base area to thickness ratio of about 7 to 30\,mm were cut from larger single crystalline parts with a diamond-wire saw. Preliminarily to the impedance measurements, the samples were coated with a thin gold layer of about 100\,nm with a Balzers Union SCD 004 sputtering device to ensure good electronic connection with the capacitor-type measurement cell. The thin side faces of the platelets were covered in wax during this process to prevent the gold layer from short circuiting the crystals. For the impedance measurements, the samples were spring-locked inside the measurement-cell, which is electrically shielded. The set-up was placed in a tube furnace under protective Ar-atmosphere and its temperature was monitored by a type-K NiCr-Ni sensor (Greisinger GTF300) placed in direct proximity to the sample. The impedance measurements were carried out with an ALPHA Dielectric Analyzer (Novocontrol Technologies) by applying a sinusoidal voltage of 0.3\,V amplitude in the range of 1\,MHz to 20\,mHz (5 points per decade). Linear voltage regime was experimentally confirmed. Impedance model fitting was done with the Z-Fit fitting routine provided in EC-Lab (BioLogic) by a Downhill-Simplex method.



\section{Data analysis and modelling}\label{modelling}

The impedance spectra of the Au coated \lmfpx\ crystals show several relaxation features visible by semi circles in Nyquist representation ($Z'$, $Z''$) or shoulders in Bode representation (${\rm{log}}(Z')$, see e.g. Fig.~\ref{IS-lfp9}). The spectra were analyzed quantitatively in terms of RQ-circuits to be able to extract the ohmic contributions of the respective features and qualitatively in the modulus formalism to distinguish between localized relaxation and long-range ionic-conduction mechanisms. The measured AC-impedance $Z$ or conductivity $\sigma _{AC} = Z^{-1}$ is related to the DC-conductivity $\sigma _{DC}$, relevant for battery applications, by a frequency ($\omega$) dependent dispersion function $F(\omega)$: $\sigma _{AC} = \sigma _{DC} \cdot F(\omega)$. The DC-conductivity can be obtained from AC-impedance data by identification of a proper dispersion function and its extrapolation to $\omega \rightarrow 0$.

Both, localized Debye-like dipole relaxations as well as ideal ionic conductivity can be described by dispersion functions characteristic for R$||$C-parallel circuits with $F(\omega) = (1 + i \omega C R)^{-1}$, where $C$ and $R$ are the respective capacitance and resistivity of the elements.~\cite{Almond1983b} This however only holds if the underlying process is characterized by a single time constant only. Real systems with finite distribution of relaxation times can be described by replacing the ideal capacitance by a constant-phase-element, parameterized by a phase angle $\alpha$ and the generalized capacitance $Q$.~\cite{Barsoukov2005} 
The resulting dispersion function reads $F(\omega) = (1 + (i \omega)^{\alpha} Q R)^{-1}$. Such distribution effects, visible by a distinct deviation of $\alpha <1$, can be particularly expected for grain boundary or surface processes.

Further information about the specific nature of a relaxation process can be obtained by analyzing the imaginary parts of the dielectric functions $Z=Z'+iZ''$ (impedance), $\epsilon=\epsilon ' - i \epsilon ''$ (permittivity), and $M=M'+iM''$ (modulus), which are related as $Z=(i \omega C_0 \epsilon)^{-1}$ and $\epsilon = M^{-1}$, where $C_0$ is the vacuum capacity of the setup. A detailed discussion of the formalism can be found in \cite{Almond1983b} and \cite{Gerhardt1994}. Following this approach, localized relaxation processes are expected to ideally show concomitant resonance peaks in the frequency dependence of $M''$ and $\epsilon ''$ caused by dielectric losses. Long-range ionic conductivity can be presumed if overlapping peaks in $M''$ and $Z''$ are found while $\epsilon ''$ stays constant.~\cite{Huggins1975,Almond1983,Almond1983b,Gerhardt1994} The resonance frequency of a long-range conduction process can be associated with a characteristic time scale, which is necessary to compensate an external electric field by migration of ionic charge-carriers inside a sample.

Distinction between local and long-range processes is not only necessary because of the occurrence of different relaxation mechanisms inside the bulk but also because of surface effects inherent to the measurement setup. Polarisation arising from the boundary layer between the sample and the ionically blocking Au-coating can be expected to superimpose the actual bulk impedance particularly at low frequencies.~\cite{Hou2010,Barsoukov2005,Karan2008} Such effects often lead to a low-frequency increase in $Z''$ and $\epsilon ''$ while the imaginary part of the modulus function $M''$ is less sensitive to surface processes~\cite{Barsoukov2005,Hodge1976}.

\section{Results and discussion}

\subsection{Anisotropic ionic conductivity in \lfp }\label{sectlfp}

\begin{figure}[htbp]
\includegraphics[width=0.5\columnwidth,clip]{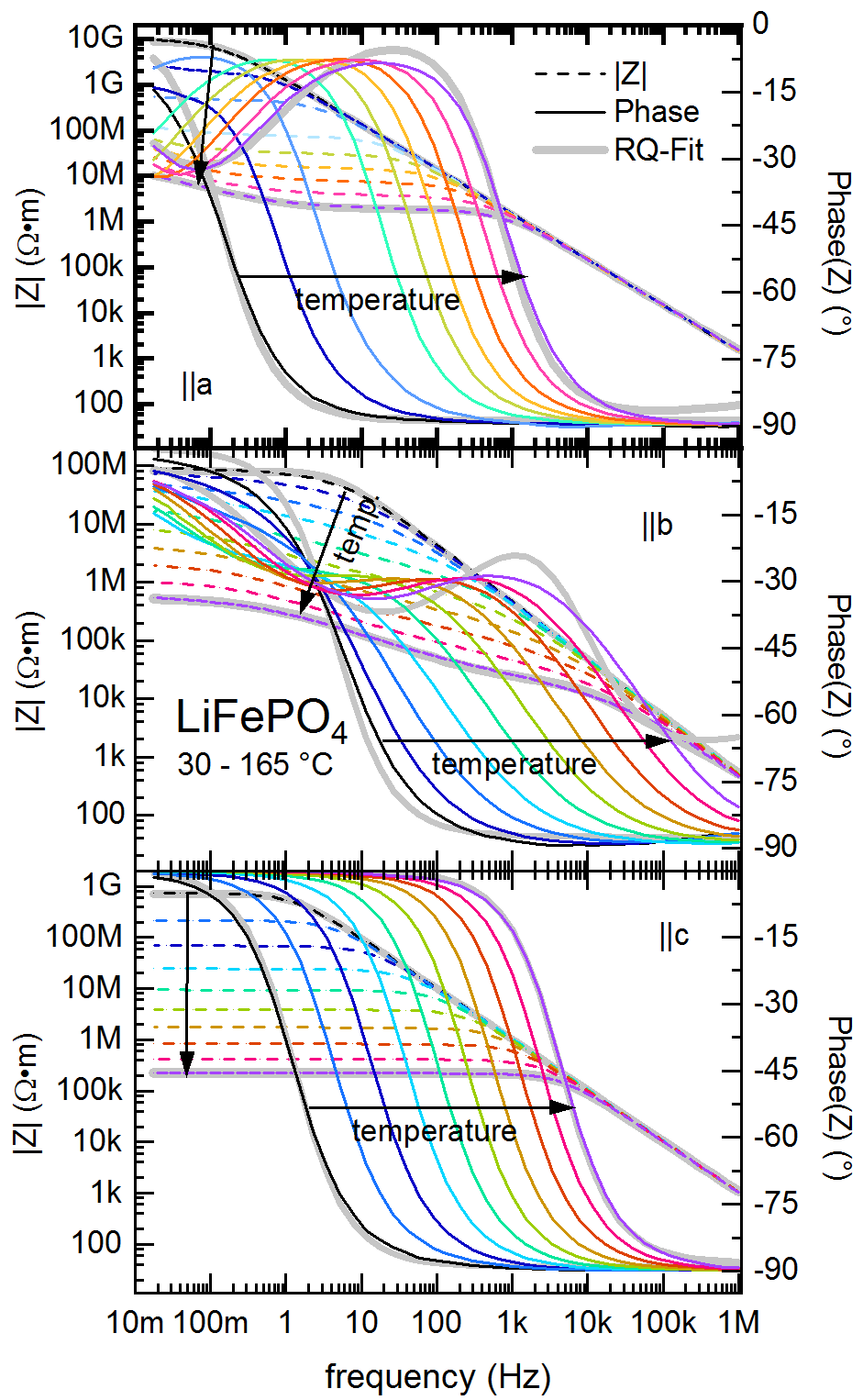}
\caption{Temperature and frequency dependence of the impedance of \lfp\ parallel to the crystallographic $a$-, $b$- and $c$-axis in Bode representation. Dashed line: absolute value of Z, solid line: phase angle of Z. The temperature variation from 30\,\celsius{} (black) to 165\,\celsius{} (purple) is color coded. Broad grey lines show fitting results as described in the text.} \label{IS-lfp9}
\end{figure}

The frequency and temperature dependance of the normalized ($\Omega$m) impedance along the crystallographic $a$-, $b$-, and $c$-directions is shown in Fig.~\ref{IS-lfp9}. For all axes, the impedance response shows a pronounced relaxation feature which is characterized by a phase change from capacitive (phase $\approx$-90\degree) to ohmic (phase $\approx$ 0\degree) behavior. Upon increasing temperature, this feature shifts to higher frequencies ($\sim$Hz at 30\,\celsius{} to $>$kHz at 165\,\celsius). For temperatures above 60\,\celsius, a second relaxation process appears at low frequencies for the $a$- and $b$-axis. As will be shown below, we attribute the high-frequency relaxation feature to long-range ionic conductivity while the low-frequency one is associated with a localized mechanism probably related to the Au-coating.

The spectra are analyzed by fitting a series of two R$||$Q-circuits describing the two features. For all spectra that do not show the low-frequency feature (e.g., at low temperatures), one of the ohmic resistors was fixed to R=0. The fit results are shown exemplarily for 30 and 165\,\celsius{} in Fig.~\ref{IS-lfp9}. The absolute values ($|Z|$) of the impedance are very well described by the fitted model. The modelled phase qualitatively describes the data. Note, that the discrepancy between measured and modelled phase of the impedance does not affect the extrapolated conductivity,  since $\sigma_{\rm DC}$ is given by the absolute value $Z$ at $\omega \rightarrow 0$.

Both relaxation features display fundamentally different properties. The feature at higher frequencies is characterized by an almost temperature independent capacitive part of the impedance of few pF. The phase angle $\alpha$ of the corresponding CPE in the model is close to unity and can be interpreted as a narrow distribution of relaxation times, i.e., rather ideal capacitive behavior. The dielectric constant of \lfp\ in this frequency regime, calculated with respect to the vacuum capacitance of the empty cell, is around $\epsilon \approx 15(5)$. The shift of the resonance frequency of this feature with temperature can thus be attributed to the temperature dependence of the bare ohmic part of Z. In contrast, the low-frequency feature shows a pronounced temperature dependance of the capacitive part along with a CPE phase-angle of about $\alpha \approx 0.55$. The latter signals strongly spread relaxation times corresponding to a pseudo capacitance of about 2\,nF, at 165\,\celsius. Considering the sample size of few mm$^2$, a capacitance of this magnitude can not be explained by polarization, but is probably caused by a chemical conversion process at the surface, e.g. Li-Au alloying in the contacting Au-layer~\cite{Barsoukov2005,Karan2008}. Since no information on the DC-conductivity can be obtained from this surface process, the low-frequency relaxation was modelled by the equivalent circuits described, but is not considered for the further analysis of the materials.

\begin{figure}[htbp]
\includegraphics [width=0.5\columnwidth,clip] {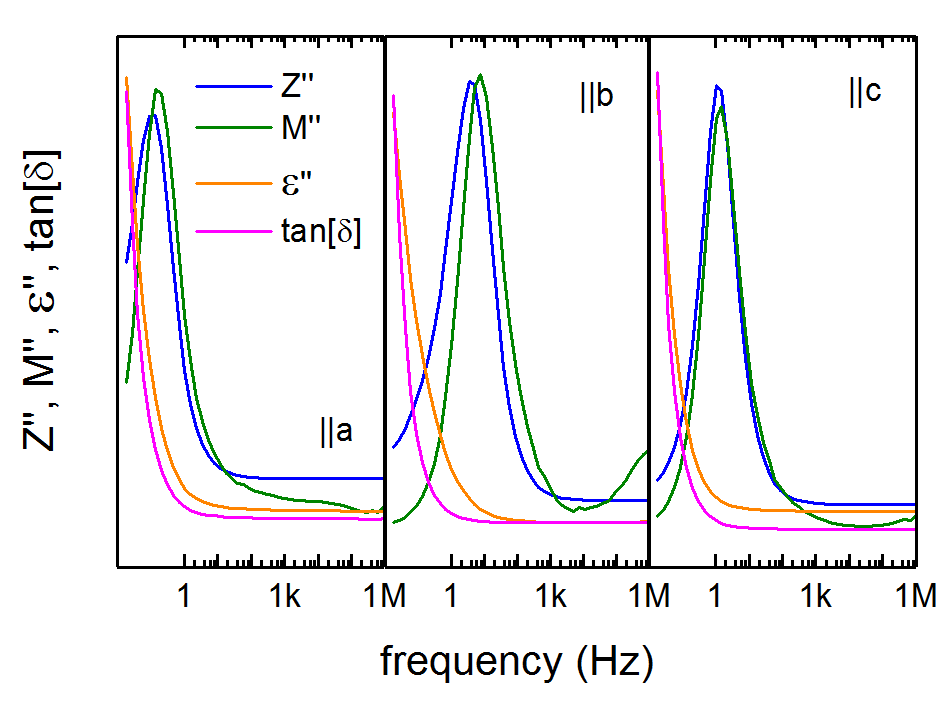}
\caption{Imaginary parts of the dielectric functions $Z$, $M$, $\epsilon$ and the loss angle $\tan(\delta)$ at 30\,\celsius{}. Concomitant maxima in $Z''$ and $M''$ indicate the long-range charge transport nature of the associated relaxation process.} \label{Mod-lfp}
\end{figure}

Relaxation at higher frequencies can be studied in the modulus formalism in more detail. Fig.~\ref{Mod-lfp} shows the imaginary part of the dielectric functions $Z''$, $M''$, and $\epsilon ''$ as well as the loss angle $\tan(\delta)$ at $T=30$\,\celsius. Along all three crystallographic axes, pronounced maxima can be found in $M''$ and $Z''$ which almost overlap in frequency. These observations strongly suggest that the process is associated with long-range charge transport (see section~\ref{modelling}). The loss angle $\tan(\delta)$ increases for lower frequencies as expected in case of finite DC-conductivity~\cite{Gerhardt1994}.

\begin{figure}[htbp]
\includegraphics [width=0.5\columnwidth,clip] {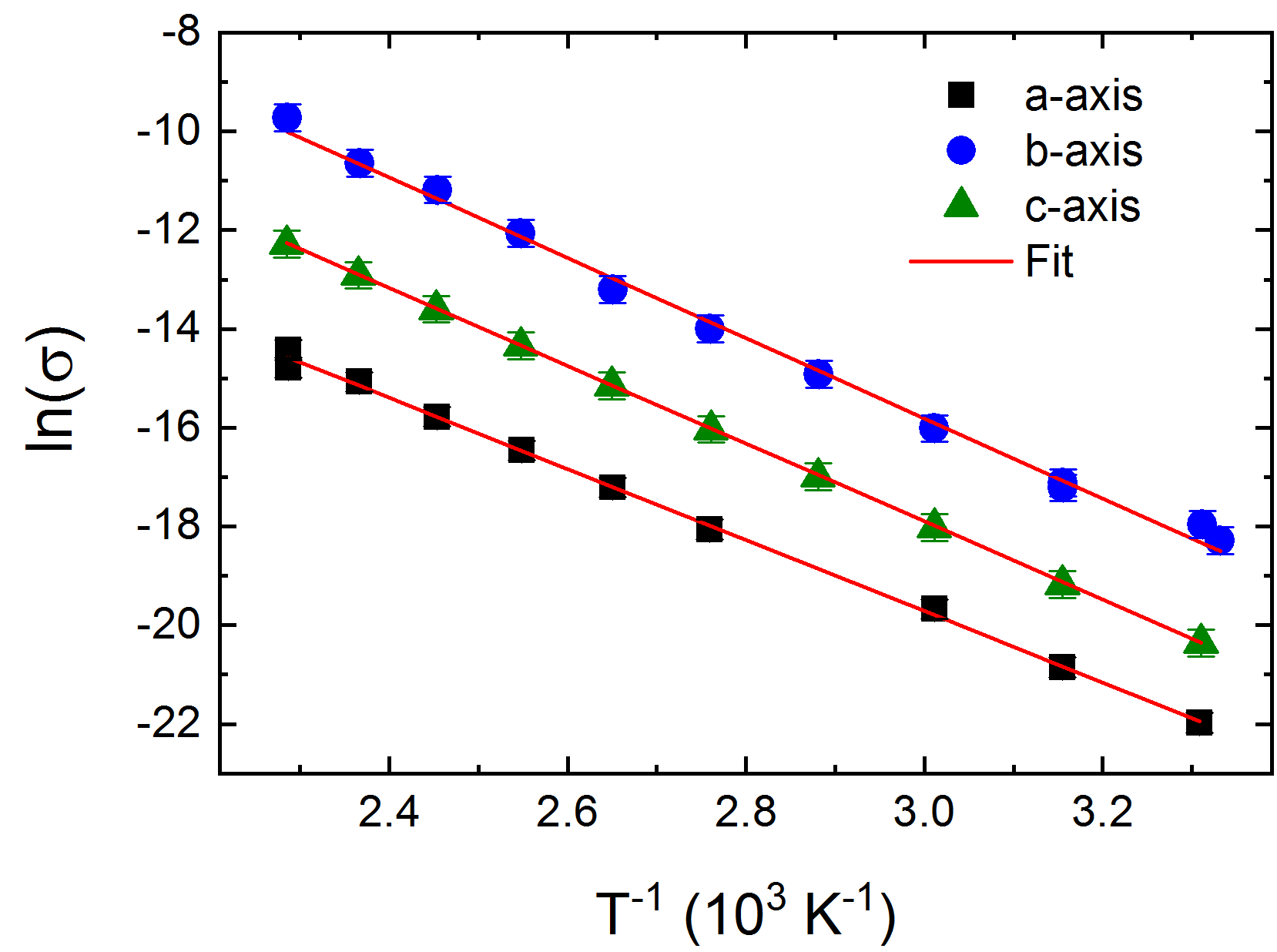}
\caption{Temperature dependence of the logarithmic ionic conductivity (data points) of \lfp\ for the different crystallographic axes and corresponding Arrhenius-fits (red lines).} \label{sig-lfp}
\end{figure}

Based on these results, one can reliably attribute the impedance response as an effect of ionic transport. In the model of R$||$Q-circuits, the DC-conductivity $\sigma _{DC}$ corresponds to the inverse resistivity $\sigma _{DC} = 1/R$. The temperature dependence of $\sigma _{DC}$ is given in Fig.~\ref{sig-lfp} shown as the logarithm of $\sigma _{DC}$ versus the inverse temperature. The data clearly show the expected activated behaviour with the $b$-axis displaying the highest conductivities. Quantitatively, the conductivities along the crystallographic $b$-, $c$- and $a$-axis differ by a factor of about 10, respectively, in the whole temperature range under study. The bulk conductivity of \lfp\ along the b-axis, at 30\,\celsius, amounts to 1.2(1)$\cdot$10$^{-10}$\,S\,cm$^{-1}$.

The linear temperature dependence as seen in the data suggests an Arrhenius-like thermal activation of the transport process. The data are fitted by functions of the form

\begin{equation}
	\sigma _{DC} = \sigma _0 \cdot \exp(-E_{\rm A}/k_{\rm B}T) \label{arr}
\end{equation}

with $E_{\rm A}$ being the activation energy of the process, $k_{\rm B}$ the Boltzmann constant and $T$ the temperature. $\sigma _0$ is a material and orientation dependent factor with minor temperature dependence. Fitting the data by means of Eq.~\ref{arr} yields the activation energies $E_A= 0.62(3)$\,eV, 0.70(3)\,eV and 0.68(2)\,eV for the $a$-, $b$- and $c$-axes, respectively. Remarkably, the $b$-direction of fast Li-transport exhibits the largest activation energy. On the other hand, although the $a$-axis is the direction of lowest conductivity, the activation energy for charge transport is slightly lower or at least comparable to the other axes. The anisotropy in the Arrhenius behaviour therefore is associated with $\sigma _0$. Considering the crystal structure and thereby the distance between ionic positions, one finds values of 3.00, 4.69 and 5.66\,\AA\ between the crystallographic Li-positions along the $b$-, $c$- and $a$-axis, respectively, which at least qualitatively implies the observed anisotropy of $\sigma _0$.~\cite{Neef2017}

The anisotropy displayed by the conductivity data in Fig.~\ref{sig-lfp} confirms the widely accepted model of 1D ionic conductivity in \lfp\ as indeed the $b$-axis conductivity provides the dominating transport channel. However, the observed anisotropy of one order of magnitude can still be regarded as moderate. In this regard it is illustrating to compare our results with previous bulk single crystal studies by Amin~\etal \cite{Amin2007} and Li~\etal \cite{Li2008}. Note, however, that owing to differences in the measurement approaches, direct quantitative comparison of measured conductivities should be restricted to the values at high temperatures which are displayed in Tab.~\ref{tabLit}.

\begin{table}[htbp]
\centering
\caption{DC-conductivity of single-crystalline bulk-\lfp , at 150\,\celsius, for the crystallographic axes as obtained in the present work and from the literature.}
\begin{tabular}{l|ccc|c}
& $a$-axis & $b$-axis & $c$-axis & \\
\hline\hline
$\sigma$ (S\,cm$^{-1}$) & 3(1)$\cdot 10^{-9}$ & 2.4(7)$\cdot 10^{-7}$ & 2.6(4)$\cdot 10^{-8}$  & present work\\
$\sigma$ (S\,cm$^{-1}$) & 1.9$\cdot 10^{-4}$  & 3.9$\cdot 10^{-4}$ & 3.7$\cdot 10^{-4}$&Ref.~\cite{Amin2007}\\
$\sigma$ (S\,cm$^{-1}$) & 2.4$\cdot 10^{-10}$ & 2.1$\cdot 10^{-7}$ & 9.0$\cdot 10^{-11}$&Ref.~\cite{Li2008} \\
\end{tabular}
\label{tabLit}
\end{table}

Note the different growth methods of the crystals used for the measurements reported in Tab.~\ref{tabLit}. The crystals studied in the present work were grown by the optical-floating zone technique at elevated Argon pressure of 30 bar and display a dark green colour.~\cite{Neef2017} Crystals used in Ref.~\cite{Amin2007} were grown in an optical furnace, too, but under ambient pressure. They display blackish green color after cleaving. Dark green crystals in Ref.~\cite{Li2008} have been grown by a standard flux growth technique. Results on the latter crystal are comparable to what was observed in the present work, however with a more pronounced anisotropy.~\cite{Li2008} Whereas, Amin~\etal\ report conductivity values which are several orders of magnitude larger than observed in the present work. It has been attributed to a mainly electronic charge transport mechanism.~\cite{Amin2007} Distinction between ionic and electronic conductivity in Ref.~\cite{Amin2007} is based on different blocking layers which are attached to the samples as well as on depolarization experiments.

\subsection{Relaxation mechanisms in \lmp }\label{sectlmp}

\begin{figure}[htb]
	\centering
		\includegraphics[width=0.5\columnwidth,clip]{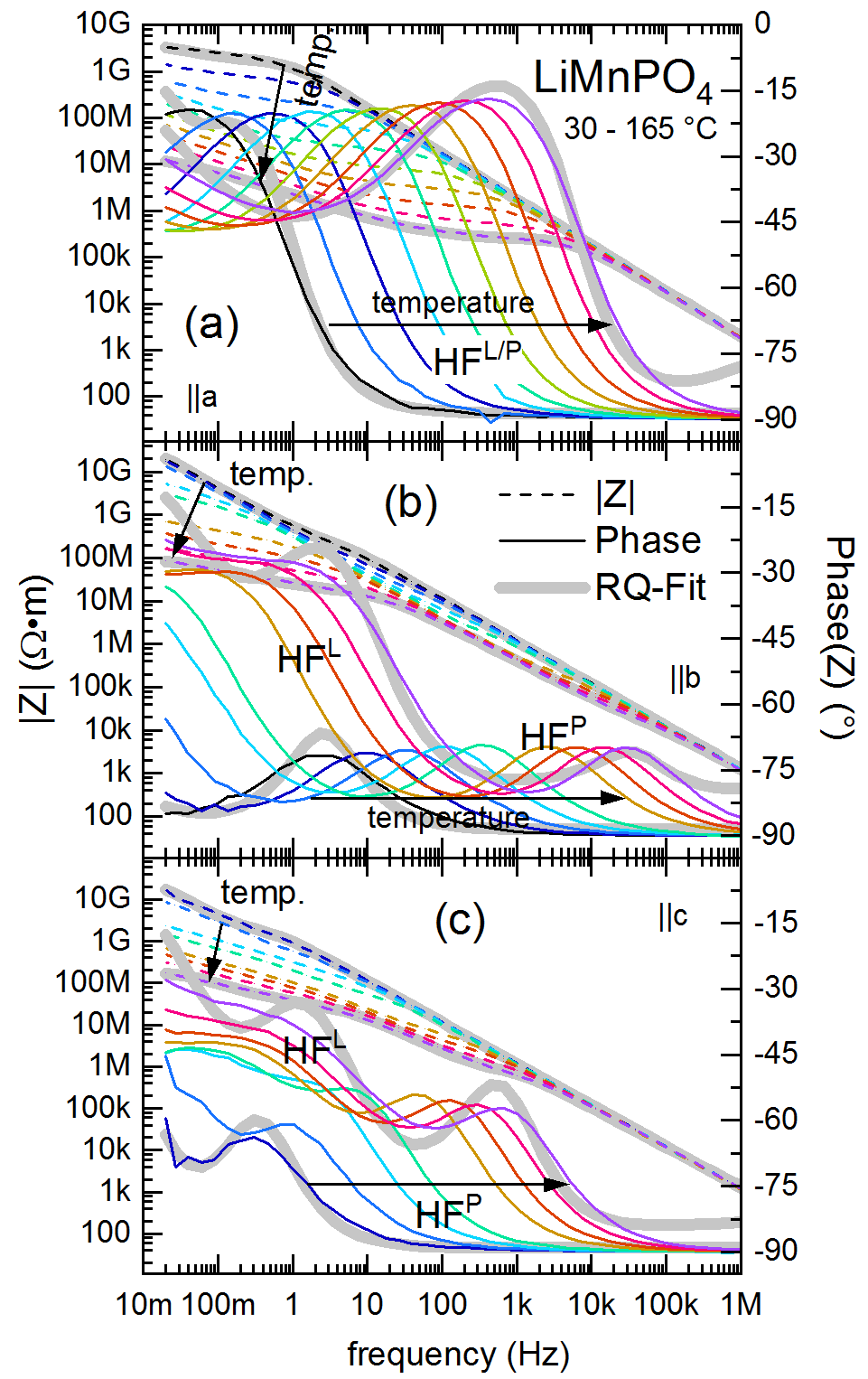}
		\caption[XXX]{Temperature and frequency dependence of the impedance of \lmp\ in Bode representation parallel to the crystallographic $a$-, $b$-, and $c$-axis. Dashed line: absolute value of Z, solid line: phase angle of Z. The temperatures from 30\,\celsius{} (black) to 165\,\celsius{} (purple) are color coded. Broad grey lines show fitting results as described in the text.}
\label{IS-lmp9}
\end{figure}

A similar approach to the interpretation of impedance spectra was carried out for \lmp . The temperature dependencies of impedance spectra along the $a$-, $b$-, and $c$-axis are presented in Fig.~\ref{IS-lmp9}. Exemplary RQ-fits are shown for 30 and 165\,\celsius{}, respectively. The data show that the impedance response of \lmp\ clearly deviates from that of \lfp . While the latter exhibits an RC-like high-frequency feature overlayed by surface effects visible at low frequencies, the response of \lmp\ shows an additional feature. Similar to \lfp , however, the low-frequency behaviour exhibits capacities up to the nF-range varying with the temperature which are associated with surface processes. The high-frequency features again correspond to capacities of a few pF.

The high-capacity surface process dominates the material's behavior already at room temperature. Along the $a$-axis, the phase shift of the impedance is smaller than -15\,\degree{} down to the lowest frequencies measured. However, while a transition towards a DC-like behavior is visible for the $a$-axis, the $b$- and $c$-axes exhibit completely different characteristics with an almost purely capacitive phase angle at low frequencies. Only at intermediate frequencies of a few Hz, a weak resonance feature (HF$^P$) is visible in the impedance at room temperature. With increasing temperature, a second resonance feature becomes visible (HF$^L$). It is associated with a pseudo-capacitance of $C_{\rm{pseudo}} \approx$~5\,$\rm{pF}$). At lowest frequencies, again the influence of surface processes becomes apparent. Within the frequency range investigated, one hence finds two different high- and mid-frequency features in \lmp\ as compared to only one in \lfp .

\begin{figure}[htbp]
	\centering
		\includegraphics[width=0.5\columnwidth,clip]{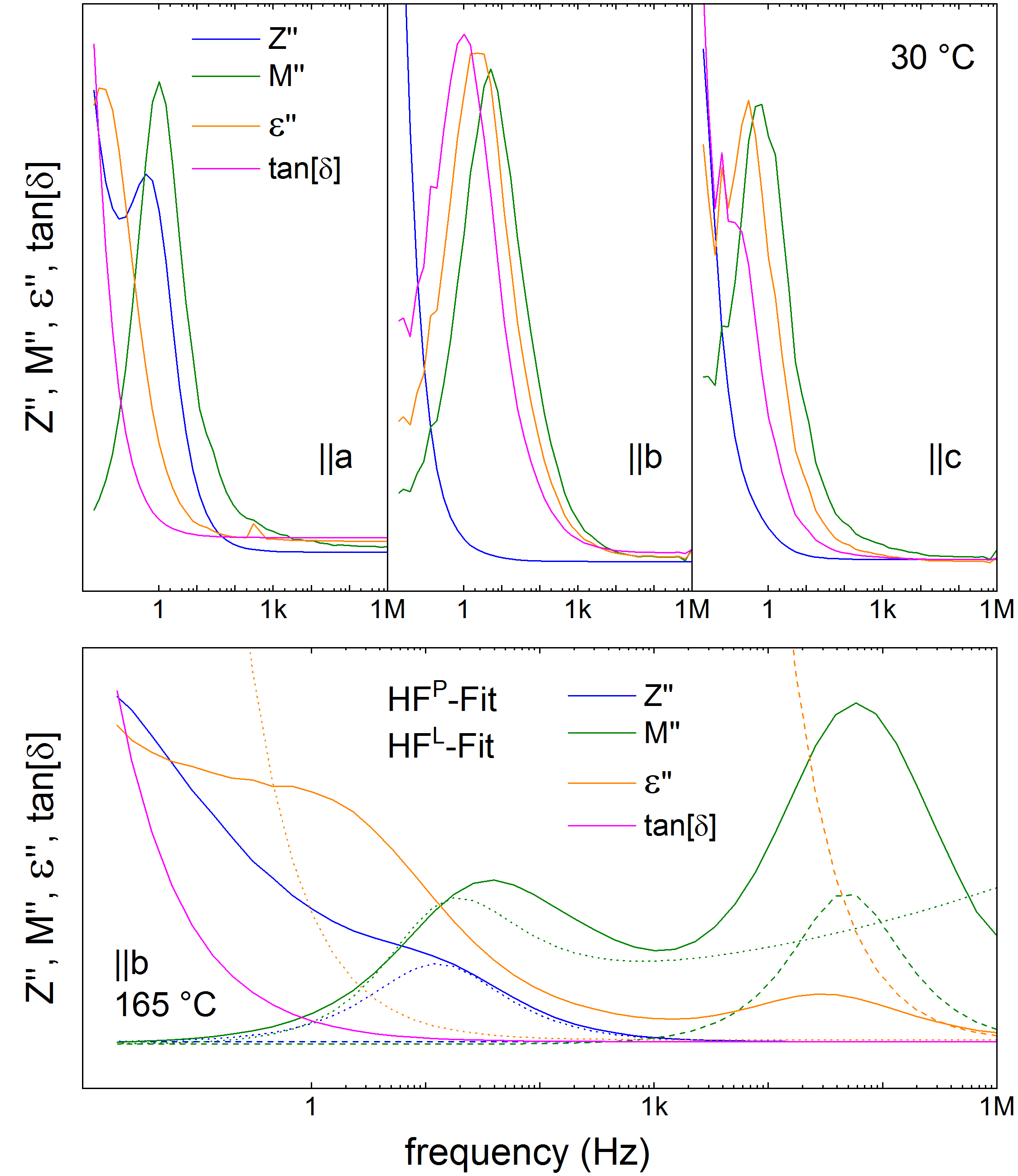}
		\caption{Imaginary part of the dielectric functions $Z$, $M$, and $\epsilon$ as well as the loss angle $\tan(\delta)$ of \lmp\ at 30\,\celsius{} (top) and 165\,\celsius{} (bottom). At 165\,\celsius{}, data measurend $||$b are shown by solid lines while dashed lines show the respective R$||$Q-fits.
 }
\label{Mod-lmp}
\end{figure}

The imaginary part of the dielectric functions along the three main axes is shown for 30\,\celsius{} in Fig.~\ref{Mod-lmp}(top). Differently to \lfp , the data do not show overlapping peaks for the dielectric functions $Z''$ and $M''$, which would identify the underlying processes to be long-range charge transport. In contrast to this, the difference of peak frequencies of $M''$, $\epsilon ''$ and $\tan(\delta)$ is rather small for the $b$- and $c$-axis response. This behavior is typical for a purely localized polarisation process~\cite{Gerhardt1994}. This conclusion is supported by the rather low dielectric relaxation ratio ($\epsilon ' (\omega \rightarrow 0) / \epsilon ' (\omega \rightarrow \infty)$) of about 4, which indicates a weak relaxation mechanism typical for an insulator. The interpretation of the $a$-axis impedances is less clear. However, it seems unlikely that there is a significant contribution of ionic long-range conductivity to the overall AC-response of \lmp\ at room temperature. This is in clear contrast to what was found for \lfp .

At higher temperatures, two different high- and mid-frequency features are visible in \lmp . The corresponding imaginary parts of the dielectric functions at 165\,\celsius{} are shown in Fig.~\ref{Mod-lmp}(bottom) for the $b$-axis. Due to the overlap of at least three different features, the frequency dependence of the dielectric functions is rather complex. Several overlapping peaks in $M''$, $Z''$ and $\tan(\delta)$ can be found. For the purpose of interpretation, the RQ-fit-functions of the two processes at high- and medium frequencies are shown in the figure. For the further analysis, the low-frequency surface process is not considered. The relaxation process marked HF$^P$, which is already visible at 30\,\celsius, shift upon heating to higher frequencies and still exhibits the characteristics of an insulator with localized polarisation. The second process, which is only visible at higher temperatures, can be identified by concomitant peaks in $Z''$ and $M''$ at frequencies between 10 to 100\,Hz. Both peaks can be assigned to the same fit-function (HF$^L$) and thus to the same relaxation process. The peak frequencies differ by about 20\,Hz. The relaxation ratio is larger than 20. The underlying process is thus likely to be of long-range nature and presumingly corresponds to ionic DC-conductivity. A clear classification, as it was feasible for \lfp , is however not possible in \lmp\ due to the difficulties of separating the individual features in the frequency domain.


 Broad distribution of relaxation times in \lmp\ is particularly challenging the data analysis. Whenever two or more relaxation processes appear on similar timescales, the sets of $R$ and $Q$ parameters, as extracted from the fit-functions, are highly correlated with each other. In the limit of two overlapping processes, only one feature will be visible in the data, which automatically leads to a misinterpretation of fit-results (e.g. $R_{\rm{fit}}=2 \cdot R_{\rm real}$, $C_{\rm{fit}}=1/2 \cdot C_{\rm real}$). In case of \lmp , one localized polarization and one long-range charge transport mechanism can be discriminated in the $b$- and $c$-axis response. One may hence speculate that the feature observed for the $a$-axis might indeed be a associated with these processes overlapping. This assumption is supported by the somewhat odd frequency dependence of the imaginary part of the dielectric functions which shows characteristics corresponding to an insulator as well as to a conductor.


\subsection{Doping dependence of the conductivity in \lmfpx }

\begin{figure}[htb]
	\centering
		\includegraphics[width=0.5\columnwidth,clip]{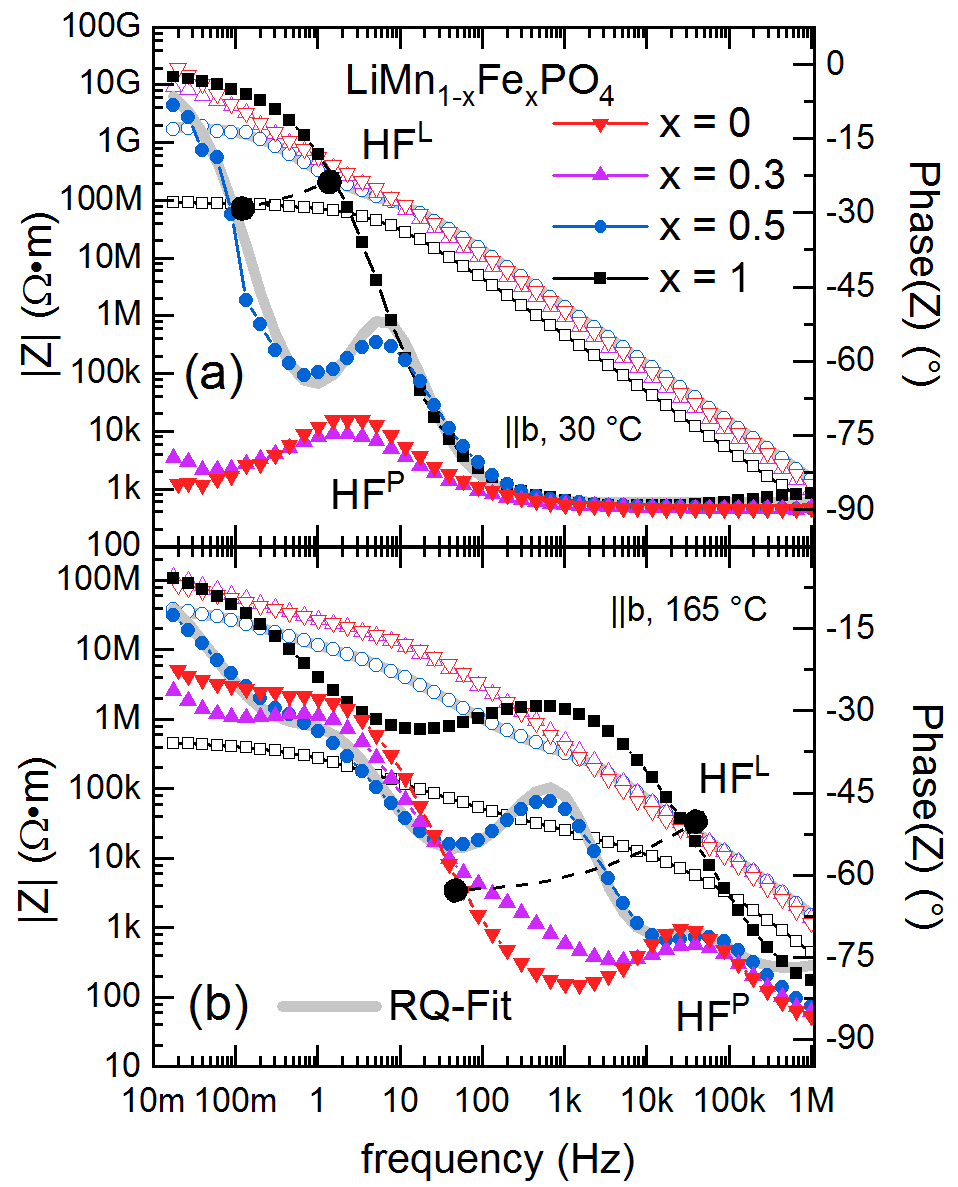}
		\caption[XXX]{Doping dependence of the impedance of \lmfpx\ at 30\,\celsius{} (a) and 165\,\celsius{} (b). $|Z|$: open symbols, phase of $Z$: filled symbols. Broad grey lines show fitting results as described in the text.}
\label{IS-all-b30165C}
\end{figure}

The end members \lfp\ and \lmp\ of the series \lmfpx\ show differences in the anisotropy of their impedance as well as in the relaxation mechanisms itself. The $b$-axis impedance responses at 30\,\celsius{} and 165\,\celsius{} of the crystals with $x=0.3$ and $x=0.5$ are shown in Fig.~\ref{IS-all-b30165C}. Compared to the end-members, the frequency dependence of the impedance of the mixed transition metal crystals shows features characteristic for both \lmp\ and \lfp . LiMn$_{0.7}$Fe$_{0.3}$PO$_4$ shows similar features as undoped \lmp . Accordingly, the behavior at 30\,\celsius{} is dominantly of capacitive nature and only a small relaxation feature (HF$^P$) can be found at frequencies of few Hz. Again, the material can be characterized as insulating without an appreciable DC-contribution to the conductivity. The impedance of LiMn$_{0.5}$Fe$_{0.5}$PO$_4$ however exhibits both, the features of a localized polarisation process as well as a change of phase-angle to a more ohmic behavior at low frequencies and thereby the characteristics of a conductor. In this respect, it is similar to \lfp . At $T=165$\,\celsius, all processes as discussed in the previous sections (i.e., high capacitance surface effects, localized polarization (HF$^P$), and presumingly long-range ionic conduction (HF$^L$)) become visible. The RQ-fit-functions are shown exemplary for both temperatures for LiMn$_{0.5}$Fe$_{0.5}$PO$_4$.

\begin{figure*}[htb]
	\centering
		\includegraphics[width=0.5\columnwidth]{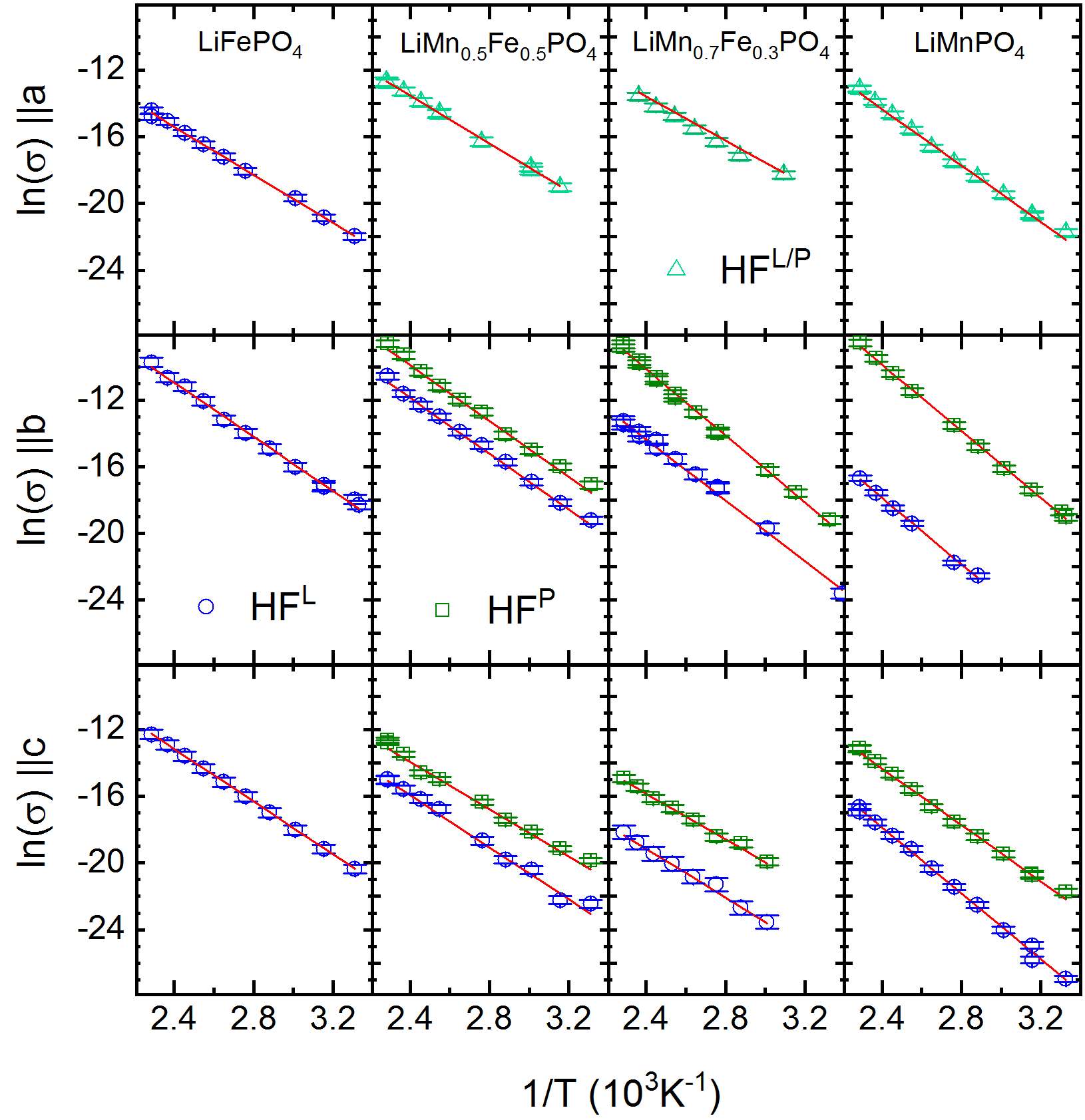}
		\caption[XXX]{Specific conductivity as extracted from the R$||$Q-Fits (blue circles: long-range conductivity, green quares: localized polarisation, turquoise triangles: unknown) and Arrhenius-fits (red lines). The temperature dependence of $\sigma$ obeys an Arrhenius-law.}
\label{Arrhenius}
\end{figure*}

All ohmic $R$-values, as extracted from the fits, were normalized with respect to the sample thickness $l$ and the surface area $A$. A compilation of the temperature dependencies of all crystals under study, converted to logarithmic specific conductivity $\log(\sigma)=-\log(R\cdot A / l)$, is presented in Fig.~\ref{Arrhenius}. Note, however, that while the dimension of a conductivity was chosen for this kind of presentation, the physical meaning of the underlying processes can only be interpreted as true DC-conductivity in the case of the long-range processes (HF$^L$). Due to the unclear distinction between long-range and localized nature of the relaxation, visible in the spectra of all Mn-containing materials along the $a$-axis, we can not conclude whether or not the $\sigma$-values $||a$ (HF$^{L/P}$) correspond to DC-conductivity (cf. section~\ref{sectlmp}).

Except for small deviations at low temperatures, the temperature dependencies of this 'effective' conductivity of all materials and along the main crystallographic directions obeys Arrhenius-like behavior. Along the $a$-axis, no clear doping dependence of the conductivity can be found. $\sigma_{\rm a}$ is of comparable magnitude for all materials with $x\neq 1$. For the $b$- and the $c$-axis, the two different relaxation mechanisms, i.e., most presumingly long-range conductivity and localized polarisation, can be well discriminated. Starting from \lmp , there is a large doping dependence of the respective effective (ohmic) conductivities. With increasing Fe content $x$, the difference between these values decreases. Specifically, the conductivity associated with ionic charge transport (HF$^L$) increases with $x$, while the effective conductivity associated with the polarization process (HF$^P$) is not strongly susceptible to the Fe/Mn ratio. Eventually, for \lfp , only one single relaxation mechanism is visible. As already mentioned above, the absence of respective evidence in the spectra does not mean that a purely polarizing mechanism in \lfp\ does not exist. It might however not be visible in AC-spectroscopy due to the comparably good ionic conductivity of the material which will dominate all competing relaxation channels. Furthermore, the resonance frequency of a respective process might be beyond the frequencies accessible in our experiment.

While different long-range ionic transport mechanisms have already been discussed in the literature for \lfp , there is no clear microscopic interpretation for the localized relaxation (HF$^P$) in the Mn-containing materials. There are however distinct differences in structural changes during Li-migration and the associated valence change of adjacent metal-ions between \lfp\ and \lmp . Li-ion or Li-vacancy migration is accompanied by redox reactions of the transition metal ions. While relatively small structural changes are expected between octahedral Fe$^{2+}$O$_6$- and Fe$^{3+}$O$_6$-groups, larger distortions can be expected in case of the MnO$_6$ octahedra due to a pronounced Jahn-Teller-effect of Mn$^{3+}$-ions.~\cite{Nie2010} It has been shown that, in \lmp , Li diffusion is strongly coupling to structural distortions of the MnPO$_4$ host which significantly affect ionic conductivity.~\cite{Rudisch2013} Li-migration is therefore not controlled by a static energy barrier between two Li-equilibrium positions, as would be caused by a static crystal potential, but strongly affected by lattice distortions. Hence, as compared to \lfp , higher activation energies can be expected for the Li-migration process. In addition, the more complex structural conditions might suggest the existence of other relaxation mechanisms that overall favor local polarization effects over long-range transport in the impedance response of the materials.

\section{Discussion and Conclusions}

\begin{table*}[htbp]
\centering
\caption{Activation energy $E_{\rm A}$ and temperature independent prefactor $\sigma_0$ (cf. Eq.~\ref{arr}) of all observed long-range relaxation processes (HF$^L$) in \lmfpx\ along the $b$- and $c$-axis as well as of the HF$^{L/P}$ process along the $a$-axis (the latter marked by an asterisk *).}
\begin{tabular}{l|ll|ll|ll}
& $\sigma _0 ||a$ & $E_{\rm A} ||a$ & $\sigma _0||b$ & $E_{\rm A}||b$ & $\sigma _0 ||c$ & $E_{\rm A}||c$ \\
& ($\Omega ^{-1}\rm{m}^{-1}$) & ($\rm{eV}$) & ($\Omega ^{-1}\rm{m}^{-1}$) & ($\rm{eV}$) & ($\Omega ^{-1}\rm{m}^{-1}$) & ($\rm{eV}$) \\
\hline
\lmp\ & 4000(800)* & 0.78(2)* & 440(270) & 0.86(3) & 330(170) & 0.85(4) \\
LiMn$_{0.7}$Fe$_{0.3}$PO$_4$ & 10(3)* & 0.57(3)* & 2100(1200) & 0.79(3) & 1(1) & 0.63(3) \\
LiMn$_{0.5}$Fe$_{0.5}$PO$_4$ & 42(10)* & 0.62(2)* & 3600(1200) & 0.72(3) & 15(10) & 0.67(3) \\
\lfp\ & 7(3) & 0.62(3) & 5000(2000) & 0.70(3) & 320(20) & 0.68(2) \\
\end{tabular}
\label{tabHFL}
\end{table*}

In order to summarize the observed doping dependence of anisotropic ionic transport in \lmfpx , the fitting parameters obtained by describing the temperature dependence of all calculated conductivities $\sigma$ associated with long-range transport (HF$^L$; HF$^{L/P}$ for the $a$-axis) by means of Arrhenius-type functions (cf. section~\ref{sectlfp}) are listed in Tab.~\ref{tabHFL}. Regarding the activation energies, two main results can be derived: Firstly, as already mentioned, the activation energy for ionic transport in \lmp\ is higher as compared to the Fe-containing materials. Secondly, for all doping levels including \lfp , the activation energy for transport parallel to the crystallographic $b$-axis is $higher$ as compared to the perpendicular directions. As discussed before, the superior ionic conductivity along the $b$-axis in \lfp\ is hence an effect of temperature independent effects absorbed in $\sigma _0$. We note, that experimental values $E_{\rm A}$ from impedance measurements sense the long-range Li diffusion and hence cannot be directly compared to results of NMR- and $\mu$SR-studies which are particularly susceptible to a short-range jump of Li-ions. This yields about 5 times smaller activation energies in the local probe experiments as compared to the results at hand.~\cite{Sugiyama}

Activation energies for Li-migration processes were calculated for \lmfpx\ with $x$ =0, 0.5, and 1 in Ref.~\cite{Gardiner2010} and for $x=0$ in Ref.~\cite{Dathar2011}. The values $E_{\rm A}$ measured in our work significantly exceed the predicted values of a Li-hopping process $Li^{\times}_{Li} \rightarrow V'_{Li}$. Dathur \etal\ report $E_A=0.29$~eV in defect-free channels along the $b$-direction. There is however a good correspondence with the ionic migration model in Ref.~\cite{Gardiner2010} and the results in Ref.~\cite{Dathar2011} considering anti-site defects. Li-\textit{M} anti-site disorder is an expected intrinsic type of defect in olivine-like transition metal phosphates.
Li-ion hopping along 1D-channels in the structure is essentially blocked by this kind of defects. Considering a concentration of about 1\,\% of displaced Li-/\textit{M}-ions would imply undisturbed 1D-channel lengths of about 30\,nm in average. The findings in Ref.\,\cite{Gardiner2010} however suggest, that while direct and long-range Li-hopping ($Li^{\times}_{Li} \rightarrow V'_{Li}$) might be suppressed, higher order Li-migration processes are possible owing to the low activation energy of anti-site defects. Hence a second 1D ($||b$-axis) transport mechanism in the presence of a disordered transition metal ion $M^{\cdot}_{Li}$ is predicted, which includes the change of position of a Li-vacancy $V'_{Li}$ and a metal ion $M^{\cdot}_{Li}$. The associated activation energy of this process ($M^{\times}_{Li} \rightarrow V'_{Li}$) is about 0.2\,$\rm{eV}$ higher as compared to Li migration ($Li^{\times}_{Li} \rightarrow V'_{Li}$). Gardiner and Islam find values of 0.7\,$\rm{eV}$ (\lfp ), 0.79\,$\rm{eV}$ (LiMn$_{0.5}$Fe$_{0.5}$PO$_4$: $Fe^{\times}_{Li} \rightarrow V'_{Li}$) and 0.87\,$\rm{eV}$ (LiMn$_{0.5}$Fe$_{0.5}$PO$_4$: $Mn^{\times}_{Li} \rightarrow V'_{Li}$) and 0.92\,$\rm{eV}$ (\lmp ).~\cite{Gardiner2010} These results are confirmed, e.g., by Dathur \etal\ who also consider anti-site defects relevant for slow diffusion in \lfp , with a barrier of 0.71~eV only for vacancy diffusion.~\cite{Dathar2011} The even quantitative correspondence to the values extracted from our measurements indicate that the long-range conductivity of the single crystalline samples used for our measurements is significantly affected by defects.

In addition to effects on 1D-transport, it is argued in Refs.~\cite{Gardiner2010,Liu2017} that the presence of anti-site defects might effectively enable ionic migration of higher dimensionality since it favours channel crossover. The motion of Li-ions, e.g., in $a$-direction from one to the next lattice position ($Li^{\times}_{Li} \rightarrow_{||a} V'_{Li}$) is energetically unfavourable. The formation of two consecutive Li-\textit{M} anti-site defects can however move a Li-ion perpendicular to the $b$-axis and might thereby offer alternative channels for effective $a$-/$c$-axis migration. The percolation threshold necessary for 3D long-range conduction was estimated to be in the range of only few \% of anti-site disordered Li-/Fe-positions for \lfp\ in Ref.~\cite{Adams2010}. With respect to defect concentrations measured in real crystals, a higher-than-one-dimensional ionic mobility cannot be excluded. Calculations in Ref.~\cite{Yang2011} even suggest that ionic migration perpendicular to the crystallographic $b$-axis might be energetically more favorable than 'overleaping' transition metal-ions residing on Li-positions ($M^{\times}_{Li}$) in $b$-direction. Following Ref.~\cite{Liu2017}, activation energy barriers along the $a$- and $c$-axes can be interpreted as the activation energy barriers for channel crossover via anti-site defects. In this respect, increase of $E_{\rm A}$ in both direction from 0.62 to 0.78\,eV for the $a$-axis and 0.68 to 0.85\,eV for the $c$-axis might reflect the decrease of the number of anti-site defects from \lfp\ to \lmp\ from about 2\% to about 1\%.~\cite{Neef2017}

\begin{figure}[htbp]
	\centering
		\includegraphics[width=0.5\columnwidth,clip]{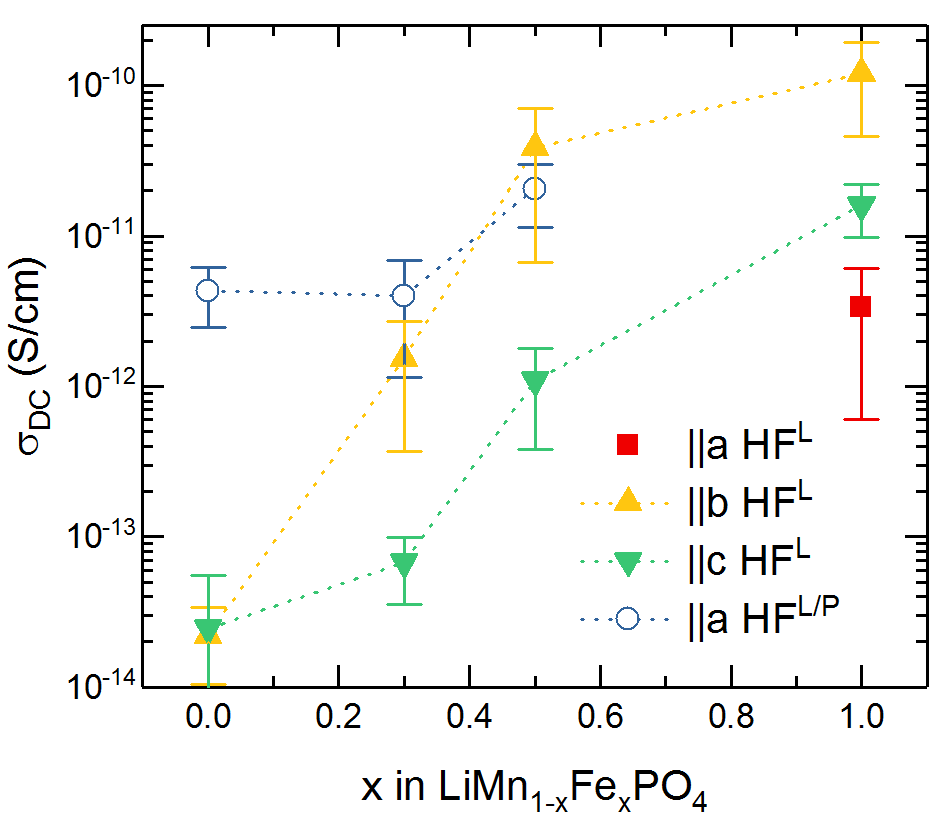}
		\caption[XXX]{Doping dependence of the ionic conductivities at 30\,\celsius{} as calculated from Arrhenius-fits.} 
\label{sigma30}
\end{figure}

Finally, Fig.~\ref{sigma30} shows the DC-conductivities related to long-range ionic conductivity, at 30\,\celsius{}, for all materials and crystallographic directions under study. In order to minimize the statistical errors, we show the conductivities as calculated by the fitted Arrhenius functions (based on 10 data points). The data illustrate the clear impact of Fe-concentration on the ionic conductivity of the materials along the $b$- and $c$-axes. The conductivity increases from $2.2(12)\cdot 10^{-14}$\,S\,cm$^{-1}$ ($||$b) and $2.4(30)\cdot 10^{-14}$\,S\,cm$^{-1}$ ($||$c) for \lmp , followed by $3.9(32)\cdot 10^{-11}$\,S\,cm$^{-1}$ ($||$b) and $1.1(7)\cdot 10^{-12}$\,S\,cm$^{-1}$ ($||$c) for LiMn$_{0.5}$Fe$_{0.5}$PO$_4$ up to $1.2(7)\cdot 10^{-10}$\,S\,cm$^{-1}$ ($||$b) and $1.6(6)\cdot 10^{-11}$\,S\,cm$^{-1}$ ($||$c) for \lfp . It is worth to highlight, that the ionic bulk conductivity of the 50\,\% Mn-doped material is of the same order of magnitude as of undoped \lfp\ along the $b$-axis, which was identified as the direction of highest conductivity of \lfp\ in this and other studies. With respect to its use in battery materials, LiMn$_{0.5}$Fe$_{0.5}$PO$_4$ would enable a high cell voltage of about 3.8\,V vs. Li/Li+ (3.4\,V for \lfp ), thereby increasing the energy density by more than 10\,\%, with comparably high Li-kinetics and hence power density.

While the anisotropy of $\sigma$ confirms quasi-1D transport in \lfp\ with fast Li-diffusion along the $b$-axis, a more complicated behaviour is found for the Mn-doped materials. As discussed above, the nature of the rather doping independent conductivity of approximately 10$^{-11}$\,S\,cm$^{-1}$ parallel to the $a$-axis (HF$^{L/P}$ relaxation) can not be unambiguously determined, i.e., whether it is associated with localized nor to long-range effects. If the observed resonance would indeed be associated to a DC-conductivity, it would however mean that the direction of highest ionic mobility changed from the $b$-axis (\lfp ) to the $ab$-plane (LiMn$_{0.5}$Fe$_{0.5}$PO$_4$), i.e., a 2D transport which is in general more tolerant to defects, and eventually to the $a$-axis (LiMn$_{0.7}$Fe$_{0.3}$PO$_4$ and \lmp ).

\section{Summary}

We report anisotropic ionic conductivity of a series of high-quality LiMn$_{1-x}$Fe$_x$PO$_4$ single crystals with doping levels 0 $\leq x \leq$ 1. Our results confirm quasi-1D transport in \lfp\ with fast Li-diffusion along the $b$-axis with $\sigma_b \approx 10\times \sigma_c \approx 100\times \sigma_a$. The anisotropy of $\sigma$ contradicts to the anisotropy of activation energies $E_{\rm A}$ of the diffusion process. The $b$-axis ionic bulk conductivity of \lmfuenf\ is of the same order of magnitude as in undoped \lfp , which implies similarly fast Li-transport for 50\% Mn-doping, i.e., an about 10~\% higher redox potential. The overall results of our impedance studies draw a far more complex picture than it would be expected from a simple 1D ionic conductor. The doping series \lmfpx\ exhibits properties partially characteristic for insulators, but also for conductors and our studies have shown the systematic appearance of localized AC-relaxation mechanisms not yet discussed in literature for this class of materials.
Regarding the mechanism of ion migration in \lmfpx , our results suggest a strong contribution of crystal defects in real materials. Particularly Li-\textit{M} anti-site defects can significantly influence the magnitude as well as the anisotropy of ionic mobility. These effects can however not be considered to be purely transport inhibiting but may also effectively increase the dimensionality of the ionic transport network.

\section*{Acknowledgments}
Financial support by the German-Egyptian Research Fund (GERF IV) through project 01DH17036, by the Deutsche Forschungsgemeinschaft (DFG) through project KL1824/5 and by the Heidelberg Graduate School for Fundamental Physics (HGSFP) is gratefully acknowledged.

\end{document}